\documentstyle[12pt]{article}

\newlength{\dinwidth}
\newlength{\dinmargin}
\begin{document}

\def\bold#1{\setbox0=\hbox{$#1$}%
     \kern-.025em\copy0\kern-\wd0
     \kern.05em\copy0\kern-\wd0
     \kern-.025em\raise.0433em\box0 }
\def\slash#1{\setbox0=\hbox{$#1$}#1\hskip-\wd0\dimen0=5pt\advance
       \dimen0 by-\ht0\advance\dimen0 by\dp0\lower0.5\dimen0\hbox
         to\wd0{\hss\sl/\/\hss}}
\def\lq{\left [}
\def\rq{\right ]}
\def\LL{{\cal L}}
\def\VV{{\cal V}}
\def\AA{{\cal A}}
\def\BB{{\cal B}}
\def\MM{{\cal M}}

\newcommand{\bea}{\begin{eqnarray}}
\newcommand{\eea}{\end{eqnarray}}
\newcommand{\nn}{\nonumber}
\newcommand{\dd}{\displaystyle}
\newcommand{\bra}[1]{\left\langle #1 \right|}
\newcommand{\ket}[1]{\left| #1 \right\rangle}
\newcommand{\spur}[1]{\not\! #1 \,}
\thispagestyle{empty}
\vspace*{1cm}
\rightline{BARI-TH/96-230}
\rightline{April 1996}
\vspace*{2cm}
\begin{center}
  \begin{Large}
  \begin{bf} 
Role of Four-Quark Operators \\
in the Inclusive $\Lambda_b$ Decays\\
  \end{bf}
  \end{Large}
  \vspace{8mm}
  \begin{large}
P. Colangelo $^{a,}$ \footnote{E-mail address:
COLANGELO@BARI.INFN.IT}  and F. De Fazio $^{a,b}$\\
  \end{large}
  \vspace{6mm}
$^{a}$ Istituto Nazionale di Fisica Nucleare, Sezione di Bari, Italy\\
  \vspace{2mm}
$^{b}$ Dipartimento di Fisica, Universit\'a 
di Bari, Italy \\

\end{center}
\begin{quotation}
\vspace*{1.5cm}
\begin{center}
  \begin{bf}
  Abstract\\
  \end{bf}
\end{center}
\noindent
We compute by QCD sum rules the  matrix elements 
of the relevant four-quark operators appearing in the expression of the 
$\Lambda_b$ inclusive decay rates at the order 
$1/m_b^3$. The results suggest that $1/m_b^3$ corrections are not responsible 
of the observed difference between the lifetime of $\Lambda_b$ and $B_d$.

\vspace*{0.5cm}
\end{quotation}

\newpage
\baselineskip=18pt
\setcounter{page}{1}
\section{Introduction}
An interesting problem of the present-day heavy quark physics is represented by 
the measured difference between the
$\Lambda_b$ baryon and $B_d$ meson lifetimes: 
$\tau(\Lambda_b^0)= 1.18 \pm 0.07 \; ps$ and
$\tau(\bar {B^0})= 1.56 \pm 0.05 \; ps$ \cite{kroll}. 
As a matter of fact, the deviation from
unity, at the level of $20 \%$, of the ratio 
$\tau(\Lambda_b)/\tau(B_d)$:
$\tau(\Lambda_b^0)/\tau(\bar {B^0}) = 0.75 \pm 0.05$,
is in contradiction with the naive 
expectation that, at the scale of the $b$ quark mass, the spectator model
should describe rather accurately the decays of the hadrons containing one 
heavy quark. The new measurement  by the Delphi Collaboration at LEP  
 of the average $b-$baryon lifetime: 
$\tau(b-baryon)= 1.25 \pm 0.11 \pm 0.05\; ps$ \cite{delphi0}, 
together with the CDF result $\tau(\Lambda_b)=1.33\pm 0.16\pm 0.07 \; ps$ 
\cite{CDF},
although pointing towards a larger value
of $\tau(\Lambda_b)/\tau(B_d)$, still confirms that this ratio
is considerably different from unity.

In principle, the ratio
$\tau(\Lambda_b)/\tau(B_d)$ can be computed in QCD. As a matter of fact,
a field theoretical approach has been 
developed for the analysis of the inclusive decay rates of the hadrons $H_Q$
containing one heavy quark $Q$ \cite{old,rev}. 
The method is based on an expansion in the inverse heavy 
quark mass $m_Q$, in the framework of the operator product expansion (OPE), and 
it has provided us with several hints on the hierarchy of the 
lifetimes of these hadronic systems.
The calculation, however, involves the matrix 
elements of a number of high dimensional quark and gluon operators.
For the $D=5$ operators defined below such matrix elements have been 
theoretically calculated, or can be inferred 
from the experimental measurements. 
As for the $D=6$ operators, in the case of heavy mesons 
their matrix elements can be obtained invoking the factorization 
ansatz, and therefore it is possible to express them in terms of quantities 
such as the leptonic constant $f_B$, 
whose estimates can be found in the literature.
On the other hand, the heavy baryon matrix elements of the
$D=6$ operators
cannot be obtained by factorization, and indeed 
they have been estimated only in the framework 
of constituent quark models, with uncertainties whose size is difficult to 
assess. Since the $D=6$ $\Lambda_b$ matrix elements  may be responsible of the 
large difference between 
$\tau(\Lambda_b)$ and  $\tau(B_d)$, 
it is interesting to compute them
by field theoretical approaches; this paper is devoted to a calculation 
based on the method of QCD sum rules. 

Before reporting on this calculation, let us briefly summarize the main
aspects of the  QCD analysis of the inclusive decay widths of the heavy 
hadrons.
The starting point is the transition operator $\hat T(Q \to X_f \to Q)$ 
\cite{old}:
\begin{equation}
\hat T=i \int d^4 x \; T[{\cal L}_W(x){\cal L}_W^\dagger(0)] 
\label{t} 
\end{equation}
\noindent 
describing an amplitude with the heavy quark $Q$
having the  same momentum in the initial and final state.
${\cal L}_W$ is the effective weak Lagrangian governing the decay
$Q \to X_f$.
The inclusive width the hadron $H_Q$ can be obtained by 
averaging $\hat T$ over $H_Q$ and taking the imaginary part of the forward 
matrix element:
\begin{equation}
\Gamma(H_Q \to X_f)={ 2 \; Im~<H_Q|\hat T|H_Q> \over 2\; M_{H_Q}} \hskip 3 pt.
\label{width} 
\end{equation}
\noindent 
The main idea to calculate the r.h.s of eq.(\ref{width})
is to set up an operator product expansion  
for the transition operator $\hat T$ in terms of local operators 
${\cal O}_i$:
\begin{equation}
\hat T= \sum_i C_i {\cal O}_i \;\;\; \label{ope}
\end{equation}
with ${\cal O}_i$ ordered according to their dimension, and the 
coefficients $C_i$ containing appropriate inverse powers of the heavy quark 
mass $m_Q$.
The lowest dimension operator appearing in (\ref{ope}) is ${\cal O}_3=\bar Q Q$.
The next gauge and Lorentz invariant operator is the $D=5$ chromomagnetic 
operator ${\cal O}_G$:
${\cal O}_G= {\bar Q} {g \over 2}  \sigma_{\mu \nu} G^{\mu \nu} Q$, whose 
hadronic matrix element
\begin{equation}
\mu_G^2(H_Q)={ <H_Q| {\bar Q} {g \over 2} \sigma_{\mu \nu} G^{\mu \nu} Q|H_Q> 
\over 2 M_{H_Q} }  \label{chro} 
\end{equation}
\noindent 
measures the coupling of the heavy quark spin to the spin of the light degrees 
of freedom in the hadron $H_Q$, and therefore 
is responsible of the mass splitting between
hadrons belonging to the same $s_\ell$ multiplet
($s_\ell$ is the total angular momentum of the light degrees of freedom in 
$H_Q$). In the case of $b$-flavoured hadrons
this mass difference has been measured,
both for mesons ($M_{B^*} - M_B = 42.0\pm \; 0.6\; MeV$ \cite{PDG}) and 
$\Sigma_b$ baryons
($M_{\Sigma^*_b} - M_{\Sigma_b} = 56 \pm 16 \; MeV$ \cite{delphi}). 

The matrix element of $\bar Q Q$ over $H_Q$ can be obtained using the heavy 
quark equation of motion, expanded in the heavy quark mass:
\begin{equation}
\bar Q Q = \bar Q \gamma^0 Q + { {\cal O}_G \over 2 m_Q^2} -
{ {\cal O}_\pi \over 2 m_Q^2} + \; O(m_Q^{-3}) \; \; \; ;
\end{equation}
${\cal O}_\pi$ is the kinetic energy operator 
${\cal O}_\pi={\bar Q} (i {\vec D})^2 Q$ whose matrix element
\begin{equation}
\mu_\pi^2(H_Q)={ <H_Q| {\bar Q} (i {\vec D})^2 Q |H_Q> \over 2 M_{H_Q} } 
\label{kin} 
\end{equation}
measures the average squared momentum of the heavy quark inside $H_Q$.
On the other hand,  the $H_Q$ matrix element  of $\bar Q \gamma^0 Q$ 
is unity
(modulo the covariant normalization of the states).

The number of independent operators appearing in (\ref{ope}) increases if the
 $1/m_Q^3$ term is considered. Such operators can be
identified in the four-quark operators of the type
\begin{equation}
{\cal O}_6^q = {\bar Q} \Gamma q \; {\bar q } \Gamma Q \label{4q}
\end{equation}
where $\Gamma$ is an appropriate combination of Dirac and color matrices.

In this way, a complete classification of the various contributions
to the inclusive decay rates can be obtained for the different hadrons $H_Q$.
In the expression for $\Gamma(H_Q \to X_f)$:
\begin{equation}
\Gamma(H_Q \to X_f)=\Gamma_0^f \; \Big[ A_0^f 
+{A_2^f \over m_Q^2} +{A_3^f \over m_Q^3} +
...\Big] \label{expan}
\end{equation}
the $A_i^f$ factors, that together with $\Gamma_0^f$ depend on the 
final state $X_f$,  
include perturbative short-distance coefficients and  
nonperturbative hadronic matrix elements incorporating the
long range dynamics.
The partonic prediction for the width in (\ref{expan})
corresponds to the leading term 
$\Gamma^{part}(H_Q \to X_f)=\Gamma_0^f A_0^f$, with $A_0^f=1+ c^f {\alpha_s 
/ \pi}+ O(\alpha_s^2)$ and $\Gamma_0^f \simeq m_Q^5$; 
differences among the widths of the  
hadrons $H_Q$  emerge at the next to leading order in $1/m_Q$ and are related 
to the different value of the matrix elements of the operators ${\cal O}_i$
of dimension larger than three.

It is important to notice the absence of 
the first order term $m_Q^{-1}$ in (\ref{expan}), 
a result obtained by
Chay, Georgi and Grinstein \cite{chay}, and Bigi, Uraltsev and Vainshtein
\cite{buv}.

The occurrence of operators of the type in eq.(\ref{4q}) 
is an appealing feature of the expansion 
(\ref{ope}), as far as the determination of the inclusive widths
is concerned. As a matter of 
fact, contrarily to the $D=5$ operators ${\cal O}_G$ and ${\cal O}_\pi$ 
which are spectator 
blind, the $D=6$ operators give different contributions when averaged over
hadrons belonging to the same $SU(3)$ light flavour multiplet, and therefore
they are responsible of the different lifetime  
of, e.g., $B^-$ and $B_s$, $\Lambda_b$ and $\Xi_b$.
The spectator flavour dependence is related to the mechanisms of weak 
scattering and Pauli interference \cite{old}, both suppressed by the factor 
$m_Q^{-3}$ with respect to the parton decay rate.

As for the differences in the lifetime of mesons and baryons, they could
already arise at the order $m_Q^{-2}$, due both to the chromomagnetic
contribution and to the kinetic energy term in (\ref{width}). 
In particular, the kinetic energy term is responsible of the 
difference for systems where the chromomagnetic contribution 
vanishes, namely in the case of $\Lambda_b$ and $\Xi_b$ having 
the light degrees of freedom in $S-$ wave. 
However, the results of a
calculation of $\mu^2_\pi$ for mesons \cite{braun}
and baryons \cite{noi} support the conjecture, put forward in \cite{wise},
that the kinetic energy operator has the same matrix element when computed on 
such hadronic systems. 
The approximate equality of the kinetic energy operator on $B_d$ 
and $\Lambda_b$ can also be inferred by  considering that,
to the leading order in $1/m_Q$, 
$\mu^2_\pi(\Lambda_b)$ can be related to  $\mu^2_\pi(B_d)$ and
to the heavy quark masses by the expression (which assumes the 
charm mass $m_c$ heavy enough for a meaningful
expansion in $1/m_c$) \cite{bigi2}: 
\begin{equation} 
\mu^2_\pi(\Lambda_b)- \mu^2_\pi(B_d) \simeq
\frac{m_b m_c}{2 (m_b-m_c)}\left[\left(M_B+3 M_{B^*}-4 M_{\Lambda_b}\right)-
\left(M_D+3 M_{D^*}-4 M_{\Lambda_c}\right)\right].\label{diff}
\end{equation}
Using present data and the CDF measurement
$M_{\Lambda_b} = 5623 \pm 5 \pm 4 \; MeV$ \cite{mlambda}
(in \cite{kroll} the value $5639 \pm 15 \; MeV$ is reported)
eq.~(\ref{diff}) gives 
$\mu^2_\pi(\Lambda_b)- \mu^2_\pi(B_d) \simeq
0.002\pm 0.024~GeV^2$, where the error mainly comes from the error on 
$M_{\Lambda_b}$.
The QCD sum rule outcome for 
$\mu^2_\pi(H_Q)$ is
$\mu^2_\pi(B_d) \simeq \mu^2_\pi(\Lambda_b) \simeq 0.6 \; GeV^2$,
with an estimated uncertainty of about $30 \%$.
This result implies that the 
differences between meson and baryon lifetimes should  
occur at the $m_Q^{-3}$ level, thus involving the four-quark operators
in eq.(\ref{4q}). They can be classified as follows  \cite{neub}:
\begin{eqnarray}
O^q_{V-A}&=&{\bar Q}_L \gamma_\mu q_L \; {\bar q}_L \gamma_\mu Q_L \nonumber \\
O^q_{S-P}&=&{\bar Q}_R q_L \; {\bar q}_L  Q_R \nonumber \\
T^q_{V-A}&=&{\bar Q}_L \gamma_\mu {\lambda^a \over 2} q_L \;
{\bar q}_L \gamma_\mu {\lambda^a \over 2} Q_L \nonumber \\
T^q_{S-P}&=&{\bar Q}_R {\lambda^a \over 2} q_L \; {\bar q}_L  
{\lambda^a \over 2} Q_R \label{4q1}
\end{eqnarray}
with $q_{R,L}= {1 \pm \gamma_5 \over 2} q$ and $\lambda_a$ the Gell-Mann 
matrices.

For mesons, the vacuum saturation approximation can be used 
to compute the matrix elements of the operators in (\ref{4q1}):
\begin{eqnarray}
<B_q | O^q_{V-A} | B_q>_{VSA} &=& \Big({m_b + m_q \over M_{B_q}} \Big)^2 
<B_q | O^q_{S-P} | B_q>_{VSA} = {f_{B_q}^2 M_{B_q}^2 \over 4}\\
<B_q | T^q_{V-A} | B_q>_{VSA} &=& 
<B_q | T^q_{S-P} | B_q>_{VSA} = 0 \;\;\; .\label{vsa}
\end{eqnarray}
Therefore, the matrix elements are expressed
in terms of quantities such as $f_B$ and the quark masses, and the 
resulting numerical 
values can be used in the calculation of the lifetimes, with the only caveat
concerning the accuracy
of the factorization approximation \cite{neub}. 

The vacuum saturation approach cannot be employed for baryons;
in this case a direct calculation of the matrix elements is required,
for example using constituent quark models. 

A simplification can be
obtained for $\Lambda_b$, 
as noticed in \cite{neub}, using color and Fierz identities and 
introducing the operators
\begin{equation}
\tilde{\cal O}^q_{V-A} = {\bar Q}^i_L \gamma_\mu Q^i_L \; 
{\bar q}^j_L \gamma^\mu q^j_L \label{otilde}
\end{equation}
and
\begin{equation}
\tilde{\cal O}^q_{S-P} = {\bar Q}^i_L  q^j_R \; 
{\bar q}^j_L Q^i_R 
\end{equation}
($i$ and $j$ are color indices). As a matter of fact, 
the $\Lambda_b$ matrix elements of the  operators in (\ref{4q1})
can be expressed in terms of 
$<\Lambda_b | \tilde{\cal O}^q_{V-A}|\Lambda_b>$ and
$<\Lambda_b | {O}^q_{V-A}|\Lambda_b>$, modulo $1/m_Q$ corrections 
contributing to subleading terms in the expression for the inclusive widths.

The matrix element of 
$\tilde {\cal O}^q_{V-A}$ and ${\cal O}^q_{V-A}$ can be parametrized as
\begin{equation}
\langle \tilde {\cal O}^q_{V-A}\rangle_{\Lambda_b} =
{<\Lambda_b | \tilde {\cal O}^q_{V-A} |\Lambda_b> \over 2 M_{\Lambda_b} } =
{f_B^2 M_B \over 48} r  \label{par}
\end{equation}
and 
\begin{equation}
{<\Lambda_b | {O}^q_{V-A} |\Lambda_b>  } = - {\tilde B} \;\;
{<\Lambda_b | \tilde {\cal O}^q_{V-A} |\Lambda_b>  } \label{btilde}
\end{equation}
with $\tilde B=1$  in the valence quark approximation.

For $f_B=200 \; MeV$ and $r=1$ eq.(\ref{par}) corresponds to the value: 
$\langle \tilde {\cal O}^q_{V-A}\rangle_{\Lambda_b} = 4.4 \times 10^{-3} \;
GeV^3$.
In ref.\cite{bilic} the $\Lambda_c$ matrix element of
$\tilde {\cal O}^q_{V-A}$ has been computed using a bag model and a
nonrelativistic quark model;  the results
$\langle \tilde {\cal O}^q_{V-A}\rangle_{\Lambda_c} \simeq 0.75 \times 10^{-3}
\; GeV^3$ and
$\langle \tilde {\cal O}^q_{V-A}\rangle_{\Lambda_c} \simeq 2.5 \times 10^{-3} 
\; GeV^3$, correspond to
$r \simeq 0.2$ and $r \simeq 0.6$, respectively.
An analysis using the model in \cite{der} can also be found in \cite{rev}.

Larger values of the matrix elements
 have been advocated by Rosner \cite{rosner}
using the values of the mass splitting $\Sigma_b^* - \Sigma_b$ and
$\Sigma_c^* - \Sigma_c$, and assuming that the $\Lambda_b$ and $\Sigma_b$ 
wave functions are similar: 
$r \simeq 0.9 \pm 0.1$ taking 
$M^2_{\Sigma_b^*} - M^2_{\Sigma_b}=M^2_{\Sigma_c^*} - M^2_{\Sigma_c}$, 
or $r \simeq 1.8 \pm 0.5$ 
using the Delphi measurement in \cite{delphi}.

It is worth supplementing the information from constituent quark models
by estimates based on field theoretical approaches, for example
QCD sum rules. As a matter of fact, a large value of $r$, namely
$r\simeq 4\;-\;5$, 
would explain the difference between $\tau(\Lambda_b)$ and $\tau(B_d)$
\cite{neub}.
As we shall see, the application of the QCD sum rule method to the calculation 
of the matrix element of an operator of high dimension
presents a number of disadvantages; 
nevertheless, interesting and quite reliable information can be obtained for 
$<\Lambda_b | \tilde {\cal O}^q_{V-A} |\Lambda_b>$. 

\vspace{2cm}

\section{QCD sum rule calculation of 
$\langle{\tilde {\cal O}^q_{V-A}}\rangle_{\Lambda_b}$}
A quantitative estimate of the matrix element 
$\langle{\tilde {\cal O}^q_{V-A}}\rangle_{\Lambda_b}$ 
can be obtained by 
the method of QCD sum rules \cite{book} applied to a suitable correlator
in the heavy quark effective theory (HQET).
Let us consider the three-point correlation function:
\begin{equation}
\Pi_{CD}(\omega, \omega^\prime )=(1+ {\spur v})_{CD} 
\Pi(\omega, \omega^\prime)
= i^2 \int dx dy <0| T [J_{C}(x) {\tilde {\cal O}^q_{V-A}}(0)
{\bar J}_{D}(y)] |0> e^{i \omega (v \cdot x)  - i \omega^\prime (v \cdot y)}
\label{threep}
\end{equation}
of the  spin $1\over 2$ local fields $J (x)$ and ${\bar J}(y)$ 
($C$ and $D$ are Dirac indices)
and of the operator $\tilde {\cal O}^q_{V-A}$ in eq.(\ref{otilde}).
The variable $\omega$ ($\omega^\prime$) is related to the residual momentum
of the incoming (outgoing) baryonic current:
\begin{equation}
p^\mu = m_b v^\mu + k^\mu
\end{equation}
with $k^\mu = \omega v^\mu$.

If the baryonic currents $J$ and ${\bar J}$ have non-vanishing
projection on the $\Lambda_b$ state
\begin{equation}
<0| J_C | \Lambda_b(v)> = f_{\Lambda_b} (\psi_v)_C \;\label{fl}
\end{equation}
($\psi_v$ is a spinor for a $\Lambda_b$ of four-velocity $v$)
with the parameter 
$f_{\Lambda_b}$ representing the coupling of the current $J$ to the 
$\Lambda_b$ state, the matrix element  
$\langle{\tilde {\cal O}^q_{V-A}}\rangle_{\Lambda_b}$ 
can be obtained by saturating the correlator (\ref{threep}) 
with baryonic states, 
and considering the double pole contribution  in the variables $\omega$ 
and  $\omega^\prime$:
\begin{equation}
\Pi^{had}(\omega, \omega^\prime) = 
\langle{\tilde {\cal O}^q_{V-A}}\rangle_{\Lambda_b}
{f^2_{\Lambda_b} \over 2} {1 \over (\Delta_{\Lambda_b} - \omega)
(\Delta_{\Lambda_b} - \omega^\prime)} 
+ \dots \label{phad}
\end{equation}
at the value  $\omega=\omega^\prime=\Delta_{\Lambda_b}$. 
The mass parameter $\Delta_{\Lambda_b}$ represents the binding energy of 
the light degrees 
of freedom in $\Lambda_b$ in the static color field generated by the $b-$quark:
\begin{equation}
M_{\Lambda_b}= m_b + \Delta_{\Lambda_b}
\end{equation}
and must be derived within the same QCD sum rule theoretical framework.

A suitable interpolating field for $\Lambda_b$, in the 
infinite heavy quark mass limit,  has been proposed by
Shuryak \cite{shuryak}.  
It is set up by a function of the quark fields:
\begin{equation}
J_C(x)= \epsilon^{i j k} ( q^{T i}(x) \Gamma \tau q^j(x))
(h_v^k)_C(x) \label{curr}
\end{equation}
where $T$ means transpose,  $i, j$ and $k$ are color indices, and
 $C$ is the Dirac index of 
the effective heavy quark field $h_v(x)$
related to the Dirac field $Q$ by:
\begin{equation}
{ h_v}(x)~=~e^{i m_Q v \cdot x} \frac{1+ {\spur v}}{2} Q(x) \; \; \; .
\label{hv}
\end{equation}
The matrix $\tau$ is 
the $\Lambda_b$ light flavour matrix:
\begin{equation} 
\tau= \frac{1}{\sqrt 2}
 \left( \begin{array}{cc} 0 & 1 \\ -1 & 0 \end{array} \right) 
\end{equation}
corresponding to zero isospin.
As implied by the spectroscopy of baryons containing 
one heavy quark in the limit $m_b \to \infty$, 
in the $\Lambda_b$ the light diquark is in a relative $0^+$ spin-parity state; 
this feature can be described by the current (\ref{curr})
by two possible choices of the Dirac matrix $\Gamma$:
\begin{equation}
\Gamma^{(1)}= C \gamma_5 \;, \; \; \; \; \; \;  
\Gamma^{(2)}= C \gamma_5 \gamma^0
\label{gamma}
\end{equation}
(for a $\Lambda_b$ at rest) where $C$ is the charge conjugation matrix. 
In principle, the currents obtained using the $\Gamma^{(\ell)}$ matrices in 
(\ref{gamma}), and a linear combination
\begin{equation}
\Gamma= C \gamma_5 ( 1 + b {\spur v} \;) \label{curr1}
\end{equation}
can be used in (\ref{threep}). As discussed in \cite{noi}, 
there are arguments in favour of the choice 
$b=1$ in (\ref{curr1}); we shall come to this point below.

Also the coupling $f_{\Lambda_b}$ can be derived by a sum rule, 
considering the two-point correlator:
\begin{equation}
H_{CD}(\omega) =
(1+ {\spur v})_{CD} H(\omega) =
i \int dx  <0| T [J_{C}(x) {\bar J}_{D}(0)] |0> e^{i \omega (v \cdot x)}
\label{twop}
\end{equation}
saturated by a set of baryonic states, in correspondence to the pole
at $\omega=\Delta_{\Lambda_b}$:
\begin{equation}
H^{had}(\omega) = {f^2_{\Lambda_b} \over 2} 
{1 \over (\Delta_{\Lambda_b} - \omega)} + \dots \;.
\end{equation}
Therefore, the calculation of $f_{\Lambda_b}$ can be carried out following the
same QCD sum rule approach in HQET; this analysis can be found in the 
literature \cite{noi,grozin}.

Let us consider the correlators (\ref{threep}) and (\ref{twop}).
In the Euclidean region, for negative values of $\omega,  \omega^\prime$ 
the correlation functions (\ref{threep})
and (\ref{twop}) can be computed in QCD, in terms of a 
perturbative contribution and of vacuum condensates.
The results can be written in a  dispersive form:
\begin{equation}
\Pi^{OPE}(\omega, \omega^\prime)= \int d\sigma d\sigma^\prime 
{\rho_\Pi(\sigma, \sigma^\prime) \over(  \sigma - \omega ) (\sigma^\prime
- \omega^\prime)} \label{pope}
\end{equation}
\begin{equation}
H^{OPE}(\omega)= \int d\sigma {\rho_H(\sigma) \over(\sigma - \omega) } \;
\end{equation}
where possible subtraction terms have been omitted.
The spectral function of (\ref{threep}) read in HQET as
\begin{eqnarray}
\rho_\Pi(\sigma, \sigma')&=&\rho^{(pert)}_\Pi(\sigma,\sigma')
+ \rho^{(D=3)}_\Pi(\sigma,\sigma') <{\bar q} q> \nonumber 
+ \rho^{(D=4)}_\Pi(\sigma,\sigma') < {\alpha_s \over \pi} G^2> \\
&+& \rho^{(D=5)}_\Pi(\sigma,\sigma') <{\bar q}g \sigma G  q> 
+ \rho^{(D=6)}_\Pi(\sigma,\sigma') (<{\bar q} q>)^2 + \dots \;\; ;\label{specf}
\end{eqnarray}
a similar expression can be given for $\rho_H(\sigma)$.

At the lowest order in $\alpha_s$ the diagrams contributing to 
$\rho_\Pi(\sigma, \sigma^\prime)$ are depicted in fig.1.
The perturbative contribution to the spectral function,
obtained computing by the Cutkosky rule the imaginary part of 
the diagram in fig.1a, has the following expression:
\begin{equation}
\rho^{pert}_\Pi(\sigma, \sigma^\prime) =
{3 \over 32 \pi^6} (1 + b^2) \big \{ \theta(\sigma- \sigma^\prime)
\sigma^{\prime  5} \; ({\sigma^{\prime  2} \over 105} - 
{\sigma \sigma^{\prime} \over 30} +
{\sigma^2\over 30} ) + (\sigma \leftrightarrow \sigma^\prime)
\big \} \; \; . 
\end{equation}
A comment is in order. The operator $\tilde {\cal O}^q_{V-A}$ in the correlator
(\ref{threep})
could give rise to non-spectator contributions through diagrams where 
two of the quark fields appearing in it 
are contracted in a tadpole. Such contributions, as noticed
in \cite{neub}, do not affect the differences in lifetime for the various heavy 
hadrons, therefore they can be omitted assuming a normal ordering in the 
four-quark operators. Such contributions vanish when 
the non-perturbative spectral functions are computed.

The contributions
proportional to the quark condensate and to the mixed quark-gluon
condensate can be derived by computing a class of diagrams of the type 
in fig.1b-d, with the result:
\begin{eqnarray}
\rho^{(D=3)}_\Pi(\sigma, \sigma^\prime)&=&
- {b \over 16 \pi^4} \big \{ 
\theta(\sigma- \sigma^\prime) \sigma^{\prime 2} (\sigma - \sigma^\prime)^2  +
\theta(\sigma^\prime - \sigma) \sigma^2 (\sigma - \sigma^\prime)^2  +
\sigma^2 \sigma^{\prime 2}  \big \} \\
\rho^{(D=5)}_\Pi(\sigma, \sigma^\prime)&=& {b \over 256 \pi^4} 
\big \{ 
\theta(\sigma- \sigma^\prime) \big(\sigma^2 +9 \sigma^{\prime 2} -
10 \sigma \sigma^\prime\big) \nonumber \\  
&+&\theta(\sigma^\prime-\sigma) \big(\sigma^{\prime 2} +9 \sigma^2 -
10 \sigma \sigma^\prime\big)  +
3 \sigma^2 +3 \sigma^{\prime 2} + 8 \sigma \sigma^\prime \big\} \;\;\; .
\end{eqnarray}
Notice that in the adopted Fock-Schwinger gauge no gluon can be emitted from 
heavy quark leg in the infinite heavy quark mass limit.

Terms proportional to the four-quark condensate come from diagrams 
of the type in fig.1e, assuming factorization of the matrix element
$<\bar q \bar q q q>$:

\begin{equation}
\rho^{(D=6)}_\Pi(\sigma, \sigma^\prime)= {1 + b^2  \over 96 \pi^2} 
\big \{\sigma^{\prime 2} \delta(\sigma) +\sigma^2 \delta(\sigma^\prime)
+\sigma \sigma^\prime \delta(\sigma-\sigma^\prime) \big \} \;\; .
\end{equation}
We do not include the gluon condensate contribution
$\rho^{(D=4)}_\Pi(\sigma, \sigma^\prime)$
since it represents the low momentum component of gluon exchange diagrams
omitted in the calculation of the perturbative term.

The inclusion of additional contributions 
proceeds in a similar way: we shall discuss in the following the consequences 
of the neglect of such terms.

A sum rule for $\langle \tilde {\cal O}^q_{V-A}\rangle_{\Lambda_b}$ 
can be derived by equating,
according to the QCD sum rule strategy, the hadronic and 
the OPE representations of the correlator (\ref{threep}).
Moreover, invoking a global duality ansatz,
the contribution of the higher resonances and of the continuum in 
$\Pi^{had}$ in (\ref{phad}) can be  modeled as the QCD contribution 
outside the region (duality region) 
$0 \le \omega \le \omega_c$,
$0 \le \omega^\prime \le \omega_c$, with $\omega_c$ an effective threshold.
Finally, the application of a double Borel transform  
to both the $\Pi^{OPE}$ and $\Pi^{had}$ representation of the correlator
(\ref{threep}) in the momenta $\omega, \omega^\prime$:
\begin{equation} 
{\cal B}(E_1)\frac{1}{\sigma-\omega}=\frac{1}{E_1} \; e^{-\sigma/E_1},\qquad
{\cal B}(E_1)\frac{1}{\Delta-\omega}=\frac{1}{E_1} \; e^{-\Delta/E_1}, 
\label{borel}
\end{equation}
(and similar for $\omega^\prime$ with the Borel parameter $E_2$) 
allows us to remove the subtraction terms appearing
in (\ref{pope}), that are polynomials in the variables $\omega$
or $\omega^\prime$ ( the Borel transform of a polynomial vanishes). 
Moreover,
the convergence of the OPE is factorially improved by the transform, and
the contribution of the low-lying resonances in $\Pi^{had}$ is enhanced for
low values of the Borel variables.
The symmetry of the spectral 
functions in $\sigma$, $\sigma^\prime$ suggests the choice $E_1=E_2=2 E$
where $E$ is the Borel parameter appearing in the QCD sum rule analysis of the 
two-point function (\ref{twop}). 
The final expression for the matrix element reads:
\begin{equation}
{f^2_{\Lambda_b} \over 2} (1 + b)^2 \exp(- {\Delta_{\Lambda_b} \over E}) 
\langle \tilde {\cal O}^q_{V-A}\rangle_{\Lambda_b} =
\int_0^{\omega_c} \int_0^{\omega_c} d \sigma d \sigma^\prime
\exp(-{ \sigma + \sigma^\prime \over 2 E} ) 
\rho_\Pi(\sigma, \sigma^\prime)
\;\;\; .\label{sr}
\end{equation}
We use the standard values  of
the condensates appearing in (\ref{specf}):  
$<\bar q q>=(- 240 \; MeV)^3$ and
$<\bar q g \sigma G q>= m_0^2 <\bar q q>$, with $m_0^2=0.8 \; GeV^2$
\cite{book}
\footnote{ In \cite{noi} the value 
$<\bar q q>=(- 230 \; MeV)^3$ has been used. The numerical results for
$f_{\Lambda_b}$ and for $\Delta_{\Lambda_b}$, within the quoted uncertainty, 
are not affected by this choice.} .
The threshold parameter $\omega_c$ has been 
fixed in the QCD sum rule determination of $f_{\Lambda_b}$ and 
$\Delta_{\Lambda_b}$ 
\cite{noi}: $\omega_c=1.1 - 1.3 \; GeV$. 
The parameter $b$ appearing in
the baryonic current $J$ in (\ref{curr})
has also been fixed in \cite{noi} studying the $\Lambda_b$ 
matrix element of the kinetic energy operator. The choice $b=1$ allowed to 
obtain
$f_{\Lambda_b}=(2.9 \pm 0.5) \times 10^{-2} \; GeV^3$ 
and $\Delta_{\Lambda_b}=0.9 \pm 0.1 \; GeV$.

Using this set of parameters, we derive from eq.(\ref{sr}) the result depicted 
in fig.2.
A stability window is observed, starting at a value of the Borel variable
$E\simeq 0.2 \; GeV$ and continuing towards large values of $E$;
in this range the result for 
$\langle \tilde {\cal O}^q_{V-A}\rangle_{\Lambda_b}$ is independent of the 
external parameter $E$.
It is known, however, that large values of the Borel variable do not provide 
us with interesting information, since in this region the sum rule 
is sensitive to the continuum model. 
Our sum rule (\ref{sr}) is particularly affected 
by this problem, due to the high dimension of the spectral density. 
Therefore, we are forced to consider a narrow region of $E$, 
close to the values $E\simeq 0.2 - 0.3 \; GeV$;
this is the Borel region already considered in the QCD sum rule
analysis of $f_{\Lambda_b}$ and $\mu^2_\pi(\Lambda_b)$ \cite{noi}.
The variation of the sum rule result with $E$ and with the continuum threshold 
$\omega_c$ provides us with an estimate of 
the accuracy of the numerical outcome. 
We find:
\begin{equation}
\langle\tilde {\cal O}^q_{V-A}\rangle_{\Lambda_b} \simeq (0.4 - 1.20) \times 
10^{-3} \; GeV^3 \;\;\; , \label{res}
\end{equation}
a result corresponding to the parameter $r$ in the range: $r\simeq 0.1~-~0.3$.

It is worth observing that the various contributions to the spectral function 
in (\ref{res}) have different signs and similar sizes, and that 
cancellations occur among the various terms. This is a common feature 
of the sum rule analyses of heavy baryon systems \cite{noi,grozin,bagan}, 
since the loop terms are 
suppressed by phase-space factors and are comparable 
with the nonperturbative 
corrections. A different procedure can be followed to soften this problem,
using a partial resummation of the nonperturbative corrections 
through nonlocal condensates \cite{rad}. Another possibility to
test of the numerical result (\ref{res}) 
consists in assuming a local quark-hadron duality prescription
\cite{rad1},
that amounts to calculate the matrix elements
of  $\langle\tilde {\cal O}^q_{V-A}\rangle_{\Lambda_b}$ and $f_{\Lambda_b}$ 
by free quark 
states produced and annihilated by the baryonic currents in (\ref{threep}) and 
(\ref{twop}), and then averaging on a duality interval in $\omega, 
\omega'$. The resulting equation for 
$\langle\tilde {\cal O}^q_{V-A}\rangle_{\Lambda_b}$ is simply given by:
\begin{equation}
\langle\tilde {\cal O}^q_{V-A}\rangle_{\Lambda_b} =
{ \int_0^{\omega_c} \int_0^{\omega_c} d \sigma d \sigma^\prime
\rho_\Pi^{pert}(\sigma, \sigma^\prime) \;
\exp(-{ \sigma + \sigma^\prime \over 2 E} ) \over 
{1 \over 20 \pi^4} 
\int_0^{\omega_c} d \sigma \sigma^5 \exp(-{ \sigma \over  E} ) }
\;\;\; ,\label{dual}
\end{equation}
where in the denominator the perturbative spectral function of the 
two-point correlator
in eq.(\ref{twop}) appears. It is possible to check the numerical
outcome for the binding energy $\Delta_{\Lambda_b}$ by this method:
the result is reported in fig.3a-b where it is shown that the same
value for such a parameter is obtained from 2 and 3-point sum rules.
As for 
$\langle\tilde {\cal O}^q_{V-A}\rangle_{\Lambda_b}$, the result is depicted in 
fig.3c; it corresponds to the value:
$\langle\tilde {\cal O}^q_{V-A}\rangle_{\Lambda_b} 
\simeq (0.6 - 1.2)  \times 10^{-3} \; GeV^3$
in agreement with the result (\ref{res}).

Let us consider, now, the parameter $\tilde B$ in eq.(\ref{btilde}).
As it emerges considering the diagrams in fig.1, in our 
computational scheme only valence quark processes are taken into account, and 
therefore a sum rule for the matrix element in (\ref{btilde}) would produce
the result $\tilde B=1$. The calculation of the contribution corresponding to 
the diagrams of the type in fig.1 immediately confirms this conclusion. 

\section{Conclusions}

Within the uncertainties of the method, we have obtained by QCD sum rules 
small values for the matrix elements of
$\langle \tilde {\cal O}^q_{V-A} \rangle_{\Lambda_b}$ and
$\langle  {\cal O}^q_{V-A} \rangle_{\Lambda_b}$, comparable with the outcome of 
constituent quark models. 
Results smaller then the value obtained using the mass splitting
$\Sigma^*_b - \Sigma_b$ suggest that an experimental confirmation of 
this mass difference is required.

From our results we conclude that the inclusion of $1/m_Q^3$ terms in the 
expression of the inclusive widths does not solve the puzzle represented by the 
difference between 
$\tau(\Lambda_b)$ and $\tau(B_d)$. As a matter of fact, 
using the formulae in \cite{neub} for the lifetime ratio, 
the value in  eq.(\ref{res}) together with $\tilde B=1$ gives:
\begin{equation}
\tau(\Lambda_b)/\tau(B_d) \ge 0.94 \; \; \;.
\end{equation}

It seems to us unlikely that the next order contribution $m_Q^{-4}$ can solve 
the problem. If the measurement of 
$\tau(\Lambda_b)$ and $\tau(B_d)$ will be confirmed in future, we feel 
that a reanalysis of the problem will be required, as 
suggested for example in \cite{alt}. Meanwhile, it is interesting that new data 
are now available for other $b-$flavoured hadrons, e.g.
$\Xi_b$ \cite{kroll}, although with errors too large to perform a meaningful
comparison with $\Lambda_b$. Such new information will be of paramount
importance for 
the complete study of the problem of the beauty hadron lifetimes, an 
argument that definitely deserves further investigation.

\vspace*{1cm}
\par\noindent
{\bf Acknowledgments}
\par\noindent
We thank V.Braun for having suggested this calculation and for discussions.
We also acknowledge interesting 
discussions with P.Ball, G.Nardulli and N.Paver.

\clearpage

\hskip 3 cm {\bf FIGURE CAPTIONS}
\vskip 1 cm

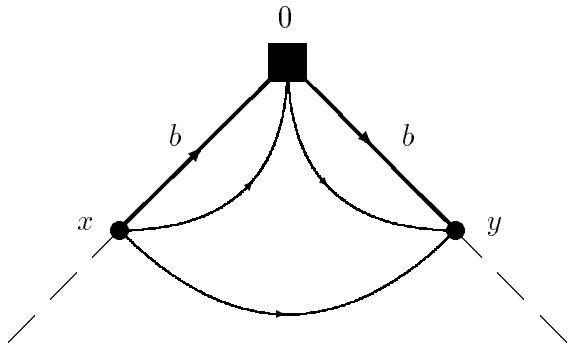
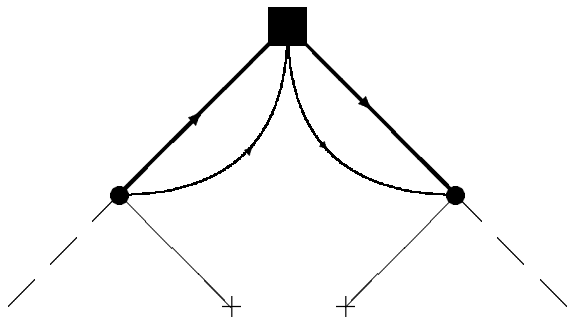
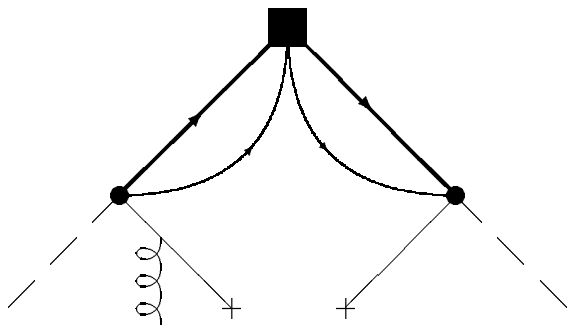
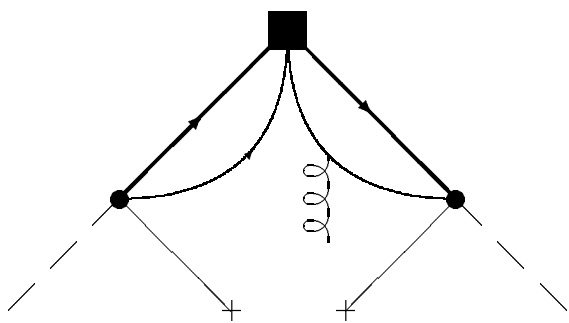
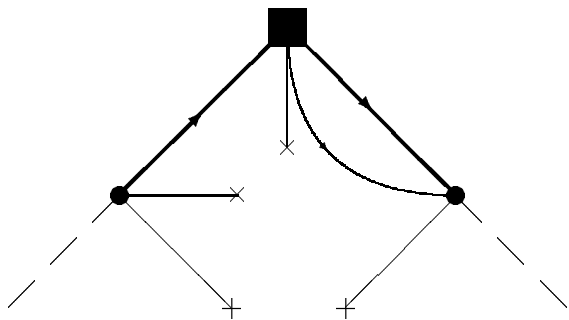
\noindent {\bf Fig. 1}\\
\noindent
Set of diagrams contributing to the spectral function (\ref{specf}).
Diagram contributing to the perturbative term (a),
to the $D=3$ term (b), to the $D=5$ term (c,d), to the $D=6$ term (e).
The thick lines correspond to the $b$-quark propagator in HQET.

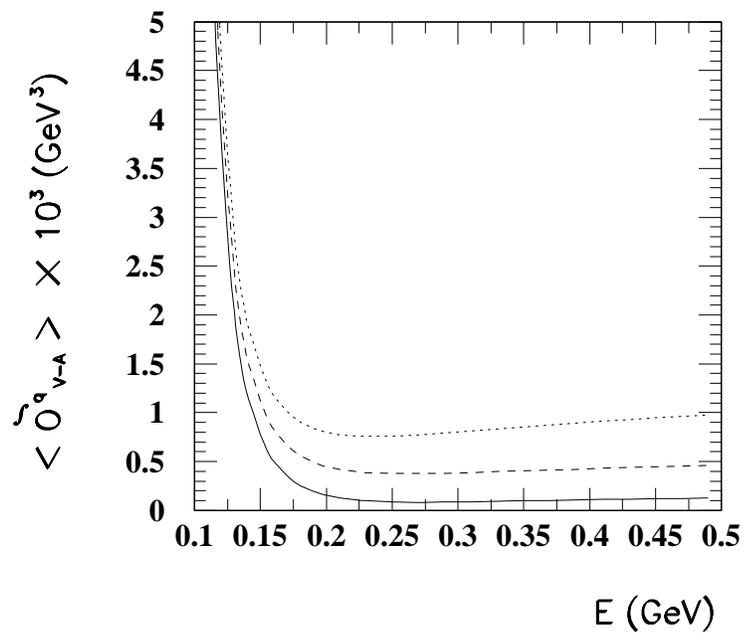
\noindent {\bf Fig. 2}\\
Sum rule (\ref{sr}) for the matrix element 
$\langle \tilde {\cal O}^q_{V-A}\rangle_{\Lambda_b}$ 
as a function of the Borel variable  
$E$. The curves refer to the threshold parameter 
$\omega_c= 1.1 \; GeV$ (continuous line),
$\omega_c= 1.2 \; GeV$ (dashed line),
$\omega_c= 1.3 \; GeV$ (dotted line).

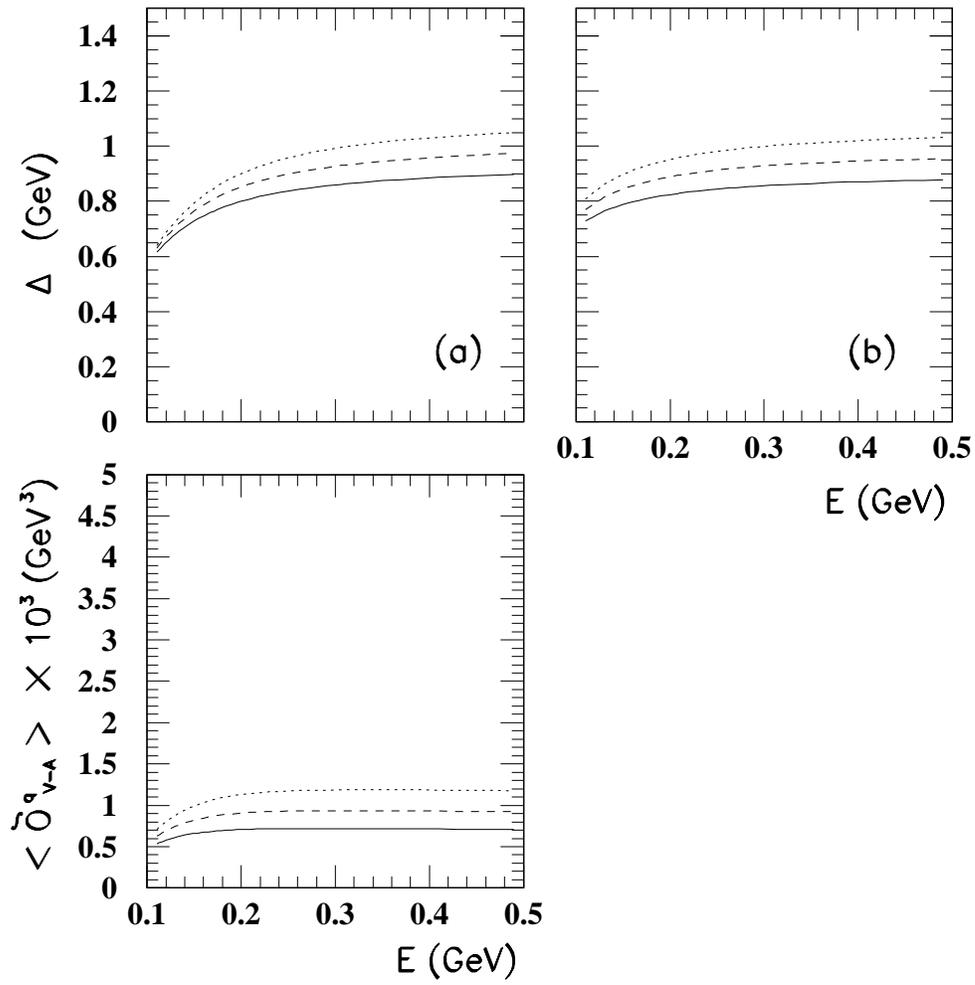
\noindent {\bf Fig. 3}\\
Sum rule (\ref{dual}) for the matrix element 
$\langle \tilde {\cal O}^q_{V-A}\rangle_{\Lambda_b}$ 
obtained using the local duality ansatz. In (a) and (b) the mass parameter
$\Delta$, obtained from the two-point sum rule and from the sum rule for
the four quark operators is depicted. 
The curves refer to $\omega_c= 1.1 \; GeV$ (continuous line),
$\omega_c= 1.2 \; GeV$ (dashed line),
$\omega_c= 1.3 \; GeV$ (dotted line).

\end{document}